\documentclass[10pt,journal]{IEEEtran}
\usepackage{cite}
\usepackage[cmex10]{amsmath}
\usepackage[ruled,vlined,linesnumbered]{algorithm2e}
\usepackage{epsfig}
\usepackage{graphicx}
\usepackage{subfigure}
\usepackage{epstopdf}
\usepackage{placeins}
\usepackage{flushend}
\usepackage{fancyhdr}
\usepackage{ifthen}
\usepackage{bm}
\usepackage{color}
\usepackage{longtable}
\usepackage{amssymb,amsmath}
\flushend
\newtheorem{definition}{Definition}

\begin{document}

\title{Designing a Disaster-resilient Network with Software Defined Networking}

\author{\IEEEauthorblockN{An Xie\IEEEauthorrefmark{1},
Xiaoliang Wang\IEEEauthorrefmark{1},
Guido Maier\IEEEauthorrefmark{2} and
Sanglu Lu\IEEEauthorrefmark{1}}

\IEEEauthorblockA{\IEEEauthorrefmark{1} National Key Laboratory for Novel Software Technology \\
Nanjing University, Nanjing, P.R. China \\
Email: waxili@nju.edu.cn}\\
\IEEEauthorblockA{\IEEEauthorrefmark{2}Dipartimento di Elettronica, Informazione e Bioingegneria\\
Politecnico di Milano,
Milano, Italy\\ Email: guido.maier@polimi.it}
}
\maketitle

\begin{abstract}
With the wide deployment of network facilities and the increasing requirement of network reliability, the disruptive event like natural disaster, power outage or malicious attack has become a non-negligible threat to the current communication network. Such disruptive event can simultaneously destroy all devices in a specific geographical area and affect many network based applications for a long time. Hence, it is essential to build disaster-resilient network for future highly survivable communication services. In this paper, we consider the problem of designing a highly resilient network through the technique of SDN (Software Defined Networking). In contrast to the conventional idea of handling all the failures on the control plane (the controller), we focus on an integrated design to mitigate disaster risks by adding some redundant functions on the data plane. Our design consists of a sub-graph based proactive protection approach on the data plane and a splicing approach at the controller for effective restoration on the control plane. Such a systematic design is implemented in the OpenFlow framework through the Mininet emulator and Nox controller. Numerical results show that our approach can achieve high robustness with low control overhead.
\end{abstract}
\IEEEpeerreviewmaketitle

\section{Introduction}

Networks having very high degree of interconnection are vulnerable to the disruptive events such as floods, earthquakes, power outages, electronic attacks, etc. Such regional damages are usually unpredictable and may simultaneously destroy multiple network facilities in a specific geographical area, which result in a long period of network outages. For example, the east Japan earthquake on March 2011 caused 385 telephone offices stopping operation immediately, cut off millions of users from the telephone service and even the emergency restoration took more than one month \cite{sakano2013disaster}.


The conventional techniques to maintain network continuity can not work well in case of disasters. Network protection, which relies on the expensive pre-allocated backup resources, may fail to deal with regional damage when the backup resources corrupt simultaneously with the primary ones. The restoration mechanism, which computes new routes based on the actual status of network, may introduce too long convergence time to meet the requirements of mission-critical and real-time applications, and leads to serious consequences like transient loops and blackholes\cite{kompella2007detection}.

To build disaster-resilient networks, this paper focuses on leveraging the technique of SDN (Software Defined Networking) which provides more intelligent and flexible network management.
 SDN networks, such as OpenFlow - enabled \cite{mckeown2008openflow} networks, decouple the network control plane from the data plane, and have been successfully deployed in the operator's WAN and corporation's LAN to provide robust network services, e.g., the global carrier NTT communications networks, Google and Microsoft inter-datacenter WANs, etc \cite{jain2013b4,hong2013achieving}. Due to its intrinsic great flexibility and global management of the network, SDN is potentially suitable to execute an efficient recovery during a major disruption.

Although SDN has a good potential for handling failures, the current architecture may be not sufficient to recover from large-scale failures such as disaster failures.
\emph{Control plane scalability} and the \emph{recovery time requirement} are the two major challenges.
Generally, the SDN controller computes routes whenever a failure occurs. And the controller is responsible for updating all the forwarding elements' status. However, the multiple failures caused by a catastrophic event will simultaneously disrupt a lot of end nodes. This will lead to a huge amount of reconnection requests, making it impractical to offload the task of all the routing computation and to update the forwarding elements' status to the controller.
This is because the dynamic route re-computation can lead to huge overhead, and inserting all new routes into SDN forwarding elements alongside is time-consuming \cite{tootoonchian2012controller} and error-prone due to the consistent packet processing problem \cite{katta2013incremental,peresini2013cpp}.
Moreover, the stringent recovery time requirements of mission-critical and real-time applications \cite{liotine2003mission} make the enhancement design of the control plane (e.g., the distributed control plane design like Onix\cite{koponen2010onix}) incompetent. This is because the status synchronization among physically distributed controllers requires additional time.

In this paper, we propose a new framework to deal with the considered problems in face of disaster failures by SDN. Several design challenges are addressed in this paper,
\begin{enumerate}
\item[(1)] \emph{Low controller overhead}. The controller overhead should be low in order to reduce the likelihood of controller being the bottleneck.
\item[(2)] \emph{Fast recovery}. The recovery should be quick in order to meet the requirement of some mission-critical and real-time applications.
\item[(3)] \emph{Strong connectivity}. The connectivity ratio should be high even after a major disaster event destroying a lot of network components.
\end{enumerate}

To address above challenges, our main idea is to pre-install redundant flow entries (backup entries) into the data plane. Different from the previous enhancement design of the control plane, our enhancement design of the data plane guarantees that a large proportion of the reconnection requests can be handled on the data plane. Since only a small fraction of the requests are handled by the controller, the control overhead is low. Besides, the data plane handled requests will not be sent to the control plane, thus saving the round-trip recovery delay between the data plane and the control plane. To address the third challenge, we consider the disaster failure's geographical layout and its failure size distribution. In order to do this, we adopt the novel metric of the vulnerable zone of a path during generating the backup entries. So the pre-installed backup routes are less likely to simultaneously get destroyed by a disaster failure. By combing with the recovery on the control plane, our design guarantees strong connectivity after a disaster failure.

More concretely, our proposed design consists of two modules: the proactive local failure recovery module running on the switches (data plane) and the reactive global restoration module running on the controller (control plane). In the protection module, we adopt the multi-topology routing to do local fast rerouting, and consider the geography properties (shape and size) of the disaster failure to generate robust backup routes. In the restoration module, we give an effective algorithm to reconnect failed nodes by rescheduling the pre-installed routes. We further consider the load balance performance during recovery by formulating an ILP, after which an heuristic algorithm is proposed. We implement the prototype by utilizing multiple tables pipeline processing and fast failover group tables of OpenFlow (Section \ref{sec:system_design}).
Simulations ( Section \ref{sec:performance_evaluation}) on both random generated and realistic topologies show that, the protection module is able to handle approximately 70\% of the reconnection requests. The rests are processed by the restoration module. Only by rescheduling the pre-installed redundancies, more than 90\% of the disconnected end nodes can be reconnected even when the failure diameter is $1/6$ of the network deployment region's length.
\section{preliminary}\label{sec:preliminary}
In this section, we first introduce the network model and failure mode adopted in this paper. Then we introduce the vulnerable zone of a routing path. \subsection{Network Model}
We consider a physical network $G(V, E)$ as a planar graph inside the deployment area $D\in \mathbb{R}^{2}$, which is represented by the network components: $V$ is the set of forwarding elements (routers or SDN switches) and $E$ is the set of links connecting them. In SDN context, all forwarding elements have a channel connected to the central \emph{controller} $C$ (in-band or out-of-band)\footnote{The logical controller \emph{C} can be implemented distributedly\cite{heller2012controller,nguyen2013software}, this refers to the controller placement problem and is out of the scope of our works.}.
By $e_{ij}$ we denote the link between adjacent nodes $i$ and $j$, $i,j\in V, e_{ij}\in E$. By $x_{st}$ we denote the path between nodes $s$ and $t$, $s,t\in V$.




\subsection{Failure Model}
During the extreme events such as disasters or malicious attacks, multiple network components located closely to each other may fail together. We summarize the behaviors of such large scale attacks to model the ``the geographically correlated failure''.


\begin{definition} (\textbf{Geographically Correlated Failure}) is defined as follows:
\begin{enumerate}
\item Network components intersecting the region of failures will be removed from the network. The size of a geographically correlated failure is determined by the radius $r$.
\item The radius $r$ follows the distribution functions $f(r)$, $r_{a}\leq r\leq r_{b}$, where $r_{a}$ (resp. $r_{b}$) is the minimum (resp. maximum) considered region size\footnote{The distribution function $f(r)$ of the destructive natural regional failures, such as earthquakes, usually follows the power-law distribution\cite{27627372}.}. 
\end{enumerate}
\end{definition}

It's notable that our model does not make any assumptions about the failure locations and radiuses, which are usually difficult to obtain due to the uncertainty of the disaster failures. Our model is more general than the previous deterministic failure model \cite{S09rbc, N08ati} (which requires the knowledge of the failure radiuses) and SRLG related model \cite{SRLG,H03dri, D04drf, S05srl, W10oao, L10dri} (which requires the knowledge of the failure locations).

\begin{figure}
	\centering
		\includegraphics[width=3.0 in]{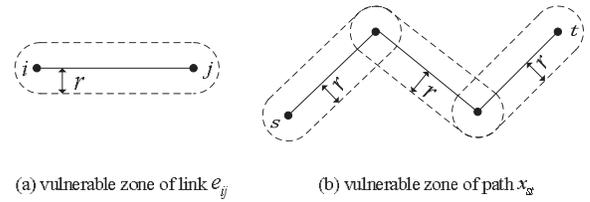}
	\caption{Vulnerable zone: the union of points that are located no more than $r$ distance from the network components. Any region failure occurs in the vulnerable zone will break the network component.}
	\label{fig:caochang}
\end{figure}
\subsection{Vulnerable Zone of a Path}
According to the definition of regional failures, a disaster region can be of any shape with arbitrary size and located anywhere in the plane. Therefore, there are infinite number of region failures to be considered. Our first problem is to find a proper statistical metric to evaluate the impact of region failures.

Given a regional failure with radius $r$, a link $e_{ij}$ may fail if it intersects with the failure region. In other words, if a disaster happens and its epicenter is less than $r$ distance from $e_{ij}$, $e_{ij}$ will be broken. We call the set of those points the \emph{vulnerable zone} of link $e_{ij}$, denoted by \textbf{$\bm{Z^{r}_{e_{ij}}}$}, defined as follows:
\begin{definition} (\textbf{Vulnerable zone of a link}) is the region sub-area such that any region failure with radius $r$  whose epicenter falls within it will always cause the corruption of the given link.
\end{definition}
As illustrated in Fig. \ref{fig:caochang} (a), the ``hippodrome'' in dash line represents the vulnerable zone of link $e_{ij}$, which consists of all points whose shortest distance to link $e_{ij}$ is no more than $r$.
Similarly, we can further define the vulnerable zone of a path $x_{st}$, denoted as $\bm{Z^{r}_{x_{st}}}$ , as shown in Fig. \ref{fig:caochang} (b), which is the union of all circle centers that are located no more than $r$ distance from the path.
\begin{definition} (\textbf{Vulnerable zone of a path}) is the region sub-area such that any region failure with radius $r$ having epicenter falling within it will always cause the corruption of the given path.
\end{definition}
The vulnerable zone of a path $x_{st}$ is the union of the vulnerable zones of all the links of the path, i.e., $Z^{r}_{x_{st}}=\cup_{e_{ij}\in x_{st}}Z^{r}_{e_{ij}}$.

\section{system design}\label{sec:system_design}
In this section, we first define the problem of SDN network reliability against regional damage. Then we introduce our system, which consists of two modules: the proactive Backup Topologies Generation Module for local recovery and the reactive Splicing Module for global restoration.

\subsection{Overview of System Design}\label{sec:overview-of-system-design}
The problem to be solved can be defined as follows: \emph{Given a network $G(V,E)$ and a central controller $C$, 1) how do we pre-install some redundancies into the network so that the controller is able to reschedule these reduncancies and 2) how does the controller reschedule the protection resources with low controller overhead to survive from the large-scale multiple failures caused by regional damage.}

\begin{figure}[!h]
\begin{center}
		\includegraphics[width=3.0in]{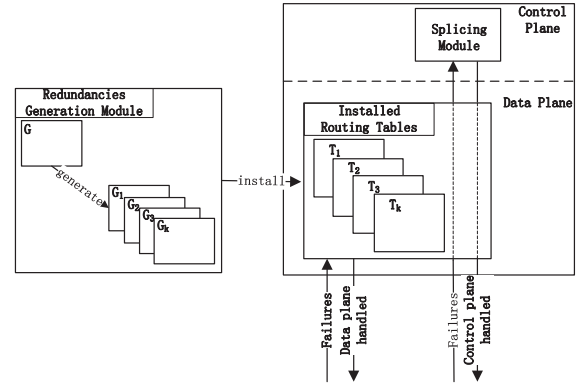}
	\caption{System overview}
	\label{fig:system_workflow}
\end{center}
\end{figure}

To solve the above problem, we apply the SDN framework for failure recovery. Our design consists of two modules, a proactive local failure recovery module working in the forwarding plane (Backup Topologies Generation Module) and a reactive global restoration module running in the control plane (Splicing Module). The failure reconnection requests first get handled by the local failure recovery module. The reconnection requests that the local recovery module is not able to handle are led to the global restoration module as illustrated in Fig. \ref{fig:system_workflow}.

How to install the redundancies for the proactive local failure recovery module needs to be carefully addressed. To reduce the controller overhead, we want to handle failures locally as much as possible. However,  the limited number of redundancies on the data plane can not handle all of the failures. We thus have to distinguish the types of failures, so as to decide which of them to handle on the data plane and which on the control plane. This ``distinction'' function can only be implemented on the data plane. Otherwise it would require the interference of the controller which may lead to some additional controller overhead (to decide the types) and additional recovery delay (the round trip time between the control plane and the data plane). To make the distinction on the data plane possible, we refer to the approach of Multi-Topology (MT) Routing (RFC 4915 and RFC 5120 \cite{psenak2007multi, przygienda2008m}) to add redundancies. And by the joint design of multi-topology redundancies and the multi-table pipeline processing of OpenFlow, this distinction function is made possible without any interference of the controller.

The basic idea of MT routing is to take the original graph $G$ as input, and generate $k$ backup topologies $\{G_1 \ldots G_k\}$. Routing tables $\{T_1 \ldots T_k\}$ are computed and installed based on $\{G_1 \ldots G_k\}$. Moreover, we notice that the recent research on MT Routing, Multiple Routing Configurations (MRC) \cite{kvalbein2009multiple,kvalbein2006fast}, is a good technique. The design goal of MRC is to prepare different configurations for different single node or link failures to achieve fast rerouting. With the adoption of MRC, the goal of the distinction function is clear, i.e., to distinguish the single link or node failure and the multiple failure. Furthermore, during the route planning in each backup topology, we consider the geographical distribution of network components to reduce the likelihood of route corruptions by regional failures.

For the reactive global recovery module running on the control plane to handle the remaining failures, a straightforward idea is to compute new routes on the controller for each failed flow and install all the new rules into the corresponding switches (\cite{staessens2011software,sharma2012openflow,nguyen2013software}). However, such an operation is time consuming and error-prone due to the consistent packet processing problem\cite{katta2013incremental,peresini2013cpp}. Instead, we exploit the usage of pre-computed backup topologies to rebuild the failed connections, and a splicing algorithm is proposed in the Splicing Module at the controller to find new paths.

\subsection{Review of MRC (Multiple Routing Configurations)}
For the completeness of our work, we first give a brief review of the MRC algorithm. The key idea of MRC routing algorithm is to prepare multiple backup topologies $\{G_1 \ldots G_k\}$, and select a proper backup topology in accordance with the current network failure state \cite{kvalbein2006fast}. In each backup topology $G_i$, some links $e_{u_{i}v_{i}}$ are defined as \emph{isolated links} and  \emph{restricted links} while some nodes $u_i$ are defined as \emph{isolated nodes}. The isolated links are set infinite weight and can be excluded from $G_i$. The restricted links are set very high weight so that they will not be chosen by some routing algorithms (i.e., shortest path routing mechanisms) unless have to. A node $u_i$ in $G_i$ is isolated if and only if its adjacent links are all either restricted or isolated.  Whenever the isolated nodes fail, it will not affect the connection of other paths.

If node $u$ detects a failure of adjacent link $e_{uv}$, $u$ will select a backup topology $G_i$ in which the failed next hop link $e_{uv}$ is isolated. Then it will tag packets with the selected backup topology id $i$ to notify the subsequent node to forward packets based on this backup topology. Since in $G_i$,  $e_{u_{i}v_{i}}$ is assigned a very high weight and it does not undertake any transit traffic, the packets are guaranteed to reach their destination.

Generally, to restore an arbitrary single link or node failure, the following constraints must be satisfied:

\begin{enumerate}
\item[(1)] Each node $u$ in the original graph must be isolated in at least one backup topology $G_i$. Each link $e_{uv}$ in the original graph must be isolated in at least one backup topology $G_i$.
\item[(2)] Each link must be isolated with one of its adjacent isolated nodes in one backup topology.
\item[(3)] All node pairs must be mutually reachable in $G_i$.
\end{enumerate}

\begin{figure}[!h]
\begin{center}
		\includegraphics[scale=0.6]{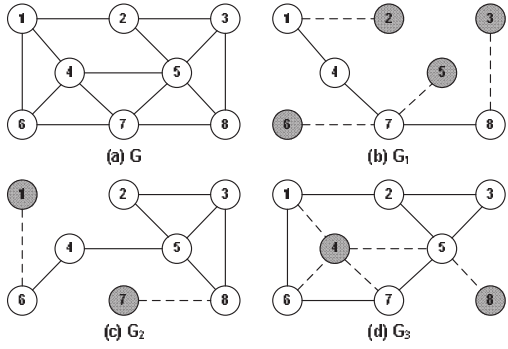}
	\caption{(a):Original network $G$.(b)-(c):backup topologies $G_1$-$G_3$. Dark node refers to the isolated node and dashed line refers to the restricted link. For clarity, we did not draw the isolated links in $G_i$.}
	\label{fig:example_backup_topologies}
\end{center}
\end{figure}

Fig. \ref{fig:example_backup_topologies} shows the generated backup topologies. Every node is isolated in exactly one backup topology. Consider a flow with (src=1,dst=3): normally its path is $1\rightarrow2\rightarrow3$ based on shortest path in $G$. Assume node 2 failed and node 1 detected the failure. Since node 2 is isolated in $G_1$, node 1 would tag the packet with tag 1 which refers to backup topology $G_1$. The alongside nodes will also forward the packet belonging to the failed flow based on $G_1$. Thus, the routing path from 1 to 3 becomes $1\rightarrow4\rightarrow7\rightarrow8\rightarrow3$.

The MRC is originally designed for locally handling single link or node failure. It is not adequate to handle multiple failures caused by large-scale regional failure. To leverage its redundancy, we first modify it based on the geographical distribution of failures to better accommodate the Splicing Module as described in the next subsection.
\subsection{Backup Topologies Generation Module}
This section first introduce how to generate routes on these backup graphs $\{G_1 \ldots G_k\}$ obtained by the MRC algorithm. Even with a sophisticated path finding algorithm, it's impossible for the limited number of redundancies on the data plane to handle all of the failures. So this module is also responsible for distinguishing the failure types so as to deliver some failures to the control plane to get handled.
\subsubsection{Geography based Backup Route Generation}
After running the MRC backup topology generation algorithm, we obtain multiple backup topologies $\{G_1 \ldots G_k\}$. If we run the same path finding algorithm (i.e., Dijkstra algorithm) on all these backup topologies, for specific nodes $s$ and $t$, the path between them, $x_{st}^{i}$ (on $G_i$) and $y_{st}^{j}$ (on $G_j$) , will be possibly coincided. This should be avoided because paths on these backup graphs are too close to each other and are vulnerable to a common risk. To improve the reliability of these paths,
we adopt the vulnerable area of  a path during the route generation in each $G_i$.

For each $s, t\in G$, we may have different paths on each $G_i$. Compared to the original path in $G$, these paths are redundant. They can be used to rebuild the failed connection between $s$ and $t$. 
However, if the vulnerable areas of backup paths intersect, they can be possibly destroyed by a regional failure simultaneously.
Consider a flow with (src=6, dst=3): normally its primary path in $G$ is $6\rightarrow 7\rightarrow 5\rightarrow 3$ based on the shortest path. The backup path from $6$ to $3$ in $G_1$ is $6\rightarrow 7\rightarrow 8\rightarrow 3$. The backup path in $G_2$ is $6\rightarrow 4\rightarrow 5\rightarrow 3$. The backup path in $G_3$ is $6\rightarrow 7\rightarrow 5\rightarrow 3$. Assume that a regional failure destroys node 5 and node 8 simultaneously. All the primary and backup paths are destroyed. This is because in this case, the primary path and the three backup paths are not region-disjoint. If, however, in $G_3$, we select the region-disjoint path $6\rightarrow 1\rightarrow 2\rightarrow 3$ from $6\rightarrow 7\rightarrow 5\rightarrow 3$ in $G_1$, backup path in $G_3$ would not get destroyed by the regional failure.



To avoid the situation in which all the primary path and backup paths are destroyed, it is required that these paths are region-disjoint\cite{T13cra}. However, finding region-disjoint paths is NP-hard even with a fixed failure radius $r$ \cite{T13cra} and is difficult to solve in general. Therefore, we refer to heuristic algorithms.
Our algorithm is shown in Algorithm \ref{algo:b_t_g}. It first finds the shortest path $x_{st}^{0}$ from $s$ to $t$ on $G$ ($G_0$) as the primary path between $s$ and $t$. Then it iterates on all the backup topologies. $[r_a, r_b]$ is evenly divided into $k$ intervals. Each backup topology $G_i$ is resilient to failures with radius up to $r_{i}=r_a+(i-1)\cdot\frac{r_b-r_a}{k-1}$. This is achieved by reducing the likelihood of the vulnerable zone of backup path, $Z_{e_{uv}}^{r_{i}}$ intersecting with the vulnerable zone of the primary path, $Z_{x_{st}^{0}}^{r_{i}}$.
If the two vulnerable zones intersect, it means that they can be both destroyed by a failure with radius $r_i$. We assign very high weight to those links whose vulnerable zones intersect with $Z_{x_{st}^{0}}^{r_{i}}$ to reduce the likelihood of choosing those links in $y_{st}^{i}$.

\begin{algorithm}[!ht]
\caption{Backup Routes Generation}
\label{algo:b_t_g}
\KwIn{network topology $G=(V, E)$, backup topologies $\{G_1 \ldots G_k\}$, source address $s$, destination address $t$}
\KwOut{$k$ backup routes in $\{G_1 \ldots G_k\}$}
\Begin{
Find $x_{st}^{0}$ in the original topology $G$\\
\For{i$\leftarrow$1 \KwTo k}{
$r_{i}:=r_a+(i-1)\cdot\frac{r_b-r_a}{k-1}$\\
\ForAll{edge $e_{uv}\in E_i$}{
\lIf{$Z_{e_{uv}}^{r_{i}}\cap Z_{x_{st}^{0}}^{r_{i}}\neq \emptyset$}{
    $w_{e_{uv}}^{i}:= very\ high\ weight$
}
}
Find $y_{st}^{i}$ in backup topology $G_i$ using the new weight
}

\Return $\{y^{1}_{st} \ldots y^{k}_{st}\}$
}
\end{algorithm}


\subsubsection{Implementation}
If the packet can not be handled on the data plane, they are sent to the controller to get handled on the control plane. The data plane should be able to deliver the failures to the control plane immediately after finding itself unable to handle them. Unlike the traditional router, the data plane and the control plane in SDN are usually physically separated. Also, the controller should not interfere with this logic. Otherwise, it would prolong the recovery time (due to the round trip time between switches and the controller) and increase the controller overhead.
We achieve this by carefully arranging the routes in the pipeline and leveraging the fast failover group table provided in the OpenFlow.

Generally, the OpenFlow pipeline processing consists of multiple routing tables  $\{T_0 \ldots T_{max}\}$ and a group table $T_g$.
Flow entries both in $T_i\ (0\leq i\leq max)$ and $T_g$ consist of a lot of terms. In $T_i$, entries consist of match fields, instructions and priority. The failover group entries in $T_g$, consist of group id and action buckets. Each action bucket is associated with a specific port (watch port) that controls the bucket's liveness. The action buckets within an entry are evaluated sequentially. The first bucket which is associated with a live port is selected.

The conventional routing procedure is to directly forward a packet $p$ to a specific port. In contrast, to leverage the fast failover group, we first forward $p$ to a specific group in the group table. Then it's up to the group to decide which port to forward to based on the port's liveness. The packets are basically divided into two types, i.e., clean packets and dirty packets. The clean packets are those packets that have not encounter any failure yet via routing. The dirty packets have encountered failures before. The clean packets and the dirty packets are processed by different processing flows in the multiple table pipeline. The two types of packets are explicitly distinguished by the MPLS tag in the packet header.

\begin{figure}[!h]
 \begin{center}
		\includegraphics[scale=0.80]{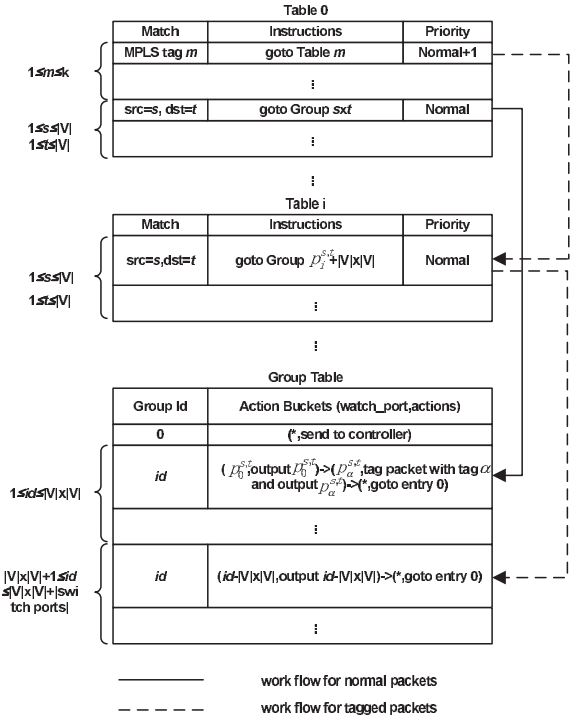}
	\caption{Prototype architecture}
	\label{fig:proposed_architecture}
 \end{center}
\end{figure}

The detailed procedure is shown in Fig. \ref{fig:proposed_architecture}. Concretely, $T_0$ is the starting table for all packets, which works as a diverting table to divert packets to different processing flows.
\begin{enumerate}
\item[(1)] A clean packet $p$ is diverted by $T_0$ to one of the groups in $T_g$. The group which $T_0$ forwards $p$ to, is responsible for checking the liveness of the output port which is based on routing table $T_0$. If the output port is alive, $p$ will be sent out via that output port. Otherwise, it will be tagged a MPLS label $\o$ ($\o$ is a backup topology's number) to indicate that it is a dirty packet. Then it will be sent out via another port which is decided by the routing in $T_{\o}$.
\item[(2)] A dirty packet $p$ with a MPLS label $i$ is diverted by $T_0$ to $T_{i}$. Then $T_{i}$ will forward $p$ to one of the groups in $T_g$. The group will check the liveness of the output port based on routing table $T_{i}$. If the port is alive, $p$ will be sent out, otherwise it will be sent to the controller to get further processed.
\end{enumerate}

\subsection{Splicing Module}\label{sec:splicing-module}
\begin{figure}
\centering
		\includegraphics[width=2.0 in]{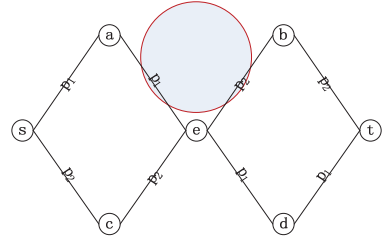}
	\caption{A regional failure destroys both $p1$ and $p2$. We rebuild the path by installing one route on node $e$ to divert traffic from $p1$ to $p2$.}
	\label{fig:splicing-motivation}
\end{figure}

The splicing module refers to the \emph{reactive} recovery at the controller. It's responsible for rebuilding the failed connections that can't not get handled by the pre-installed redundancies. As described in Section \ref{sec:overview-of-system-design}, unlike conventional approaches that install all routes into forwarding elements, we rebuild the failed connections by utilizing the pre-installed redundancies. By doing so, the number of installed routes is reduced, thus reducing the likelihood of the consistent packet processing problem. A motivation example is shown in Fig. \ref{fig:splicing-motivation}. The are two paths from $s$ to $t$, $p1$ and $p2$. A regional failure destroy $e_{ae}$ and $e_{eb}$ simultaneously. As a result, both $p1$ and $p2$ are destroyed. To rebuild the connection, the traditional approach is to install new routes along $s\rightarrow c\rightarrow e\rightarrow d\rightarrow t$, which requires installing five new rules. In contrast, we only install two rules, one on node $s$ to divert traffic to $p2$, the other one on node $e$ to divert traffic from $p2$ to $p1$.

Another issue we consider is the load balancing during recovery. Unlike the single link or node failure, the regional failures usually destroy a huge amount of network components simultaneously and lead to a huge amount of disconnected end nodes. Such a huge number of disconnected end nodes requires lots of reconnections. The splicing module should handle the reconnections in a proper way to avoid that some nodes bear exceedingly more rerouting paths than others. The requests that sent to the controller give us the opportunity to redistribute some of the reconnecting traffic. Aside from the connectivity, we also consider the problem of how to reconnect the failed paths. We first define a metric, then we give an ILP to formulate the problem, after which we propose an efficient heuristic algorithm to reduce the complexity.

We define the metric, maximal load to quantify the routing load balance degree after a regional failure.
\begin{definition} (\textbf{Maximal Load}): given a network graph $G^{'} (V^{'}, E^{'})$ after a regional failure (with failure radius $r$), and a reconnection request matrix $\mathcal{RM}$ on $G^{'}$, the maximal load gap is the load of the most loaded node in the network, where the load of a node is the number of primary and rerouted paths passing it, i.e.,
\begin{align} \nonumber
ML = |(P_{u}+R^{'}_{u})|_{max}, \forall u \in V^{'}
\end{align}
\end{definition}
By $P_{u}$ we denote the number of primary paths that pass  $u$. By $R^{'}_{u}$ we denote the number of the rerouted paths that pass $u$ based on $\mathcal{RM}$ after a regional failure. Here, we consider the primary resource and the rerouted resource separately, and we assume that if a primary path for a particular node pair is not failed, the primary path can not be altered to avoid network wide reconfiguration. Our goal is to minimize the maximal load.

We formulate the problem by the following ILP,
\begin{alignat}{1}
\min \max_{ i \in V^{'}}\quad &\bigg(\sum_{\forall s, t \in V}\sum_{e_{ij}\in E}x^{st}_{ij} + \sum_{\forall s, t \in \mathcal{RM}_{1}}\sum_{e_{ij}\in E^{'}}y^{st}_{ij}\nonumber \\ \tag{ILP1}
\!\!\!\!\!\!\!  & + \sum_{\forall s, t \in \nonumber\mathcal{RM}_{2}}\sum_{e_{ij}\in E^{'}}z^{st}_{ij}  \bigg)\\ \nonumber
\mbox{s.t.}\quad
&z_{st}\ is\ a\ routing\ path\tag{1a}\nonumber\\
&\!\!\!\!\!\!\!\! y_{ij}^{st} \in \{0, 1\}, \forall e_{ij}\in E,\forall s, t \in V^{'} \tag{1b} \nonumber
\end{alignat}

The first part in the object function corresponds to $P_{i}$, in which $x_{ij}^{st}$ equals to 1 if path $x_{st}$ passes $e_{ij}$ and 0 otherwise. $x_{ij}^{st}$ is computed based on the routing in $G_{0}$. The second part in the object function corresponds to the additional rerouting paths that $i$ have to undertake due to the reconnection requests $\mathcal{RM}_{1}$. $\mathcal{RM}_{1}$ is the reconnection requests that can be handled on the data plane. The second part can also be computed. The third part in the object function computes the number of rerouting paths that $i$ have to undertake due to the reconnection requests $\mathcal{RM}_{2}$. $\mathcal{RM}_{2}$ is the reconnection requests that can not be handled on the data plane. It is notable that $\mathcal{RM}$ equals to the sum of $\mathcal{RM}_{1}$ and $\mathcal{RM}_{2}$.

ILP1 distribute the rerouting traffic in a min-max fashion. However, the above optimization may not scale well to large networks. In addition to the optimization, we also give a heuristic algorithm. To evenly distribute the reconnections request using the installed redundancies, for all node $u\in E$, we record $R^{'}_{u}$ on the control plane. We first construct a temporary graph in which, we fill all the available segments (not broken by the regional failure) from $\{G_1 \ldots G_{k}\}$ into the temporary graph. Then weight assignment is performed based on $R^{'}_{u}$ for all node $u\in E$. The detailed algorithm is in Algorithm 2.

In Algorithm 2, we first test if $s$ and $t$ are physically disconnected. If they are, it's impossible to find a path between them. Then we construct a multigraph $G_{temp}$ by adding edges from $k$ paths from $s$ to $t$ in $\{G_1 \ldots G_k\}$ excluding the failed edges. To evenly distribute the rerouting paths, we set link weight of $e_{uv}$ to the mean of $R^{'}_{u}$ and $R^{'}_{v}$. After the weight is set, we try to find a path on this temporary graph. If a path is found, we update $R^{'}_{u}, u\in V$ and return the splicing actions. Otherwise, it means that the failed path can not be rebuild by splicing the existing redundancies. In this case, one may have to install a whole new path.
\begin{algorithm}[!ht]
\caption{Splicing Action Generation}
\label{algo:s_a_g}
\KwIn{network topology $G=(V, E)$, backup topologies $\{G_1 \ldots G_k\}$, source address $s$, destination address $t$}
\KwOut{a set of splicing actions.}
\Begin{
\eIf{$s$ and $t$ are physically disconnected}{
\Return failed and abort
}
{
build a temporal topology $G_{temp}(V_{temp}, E_{temp})$\\
$V_{temp}:=V, E_{temp}:=\emptyset$\\
\For{i$\leftarrow$1 \KwTo k}{
\ForAll{edge $e_{uv}\in E_i$}{
\eIf{$e_{uv}$ is alive and on the path from $s$ to $t$ in $G_i$}{
$E_{temp}:=E_{temp}\cup e_{uv}$\\
$\ell_{e_{uv}}:=i$
}{
continue
}
}
}
\ForAll {edge $e_{uv}\in E_{temp}$}{
$w_{e_{uv}}:=(R^{'}_{u}+R^{'}_{v})/2$
}
find shortest path $p_{temp}$ on $G_{temp}$ from $s$ to $t$\\
\eIf{found}{
\ForAll {edge $e_{uv}\in E_{temp}$}{
\lIf{$e_{uv}\in p_{temp}$}{\\
$R^{'}_{u}:=R^{'}_{u}+1$\\
$R^{'}_{v}:=R^{'}_{v}+1$
}
{}
}
}{\Return failed and abort}
\Return splicing actions based on $p_{temp}$
}
}
\end{algorithm}

%
\section{performance evaluation}\label{sec:performance_evaluation}
\subsection{Simulation Setting}
 We use both random and realistic topologies for our simulation. The random50 topology contains 50 nodes and 120 edges, the random100 topology contains 100 nodes and 211 edges. The realistic Germany backbone consists of 50 nodes and 88 edges. The deployment area is 1200 x 1200 (arbitrary units) for all the cases. One connection is requested by each node pair of the network. We implement our prototype using OpenFlow 1.3.3\cite{specification2013version} and NOX controller\cite{gude2008nox} and examine the prototype's performance on Mininet testbed\cite{lantz2010network}. For the throughput test, we use two PCs, one running mininet and the other running NOX controller. Iperf \cite{tirumala2005iperf} is adopted as our test tool.

 \textbf{Comparsion Metrics.} We use the following metrics to quantify the results\cite{motiwala2008path, wangfast}.
\begin{itemize}
\item \emph{Recovery Ratio}.
Recovery ratio is introduced to evaluate the capacity of network recovery to reset the connection between pairs of disconnected nodes, which can be defined as follows,
\begin{definition}
\textbf{(Recovery Ratio)}
$$\mbox{Recovery Ratio}= {\frac{\mbox{number of recovered paths}}{\mbox{number of recoverable paths}}}$$
\end{definition}
As a result of multiple failures, the underlying topology may be divided into disconnected components, or the source (and/or destination) of a certain flow becomes failed. Hence, we call a disconnected routing path as ``recoverable'' if both the end nodes are alive and they are not physically separated.
\item \emph{Path stretch}. The detail definition can be found in \cite{W13fcr}. Generally, a path with longer stretch requires more network resources. We adopt the notation of \emph{stretch} to measure the ratio of alternate path length over the expected shortest path length.
\item \emph{Controller overhead}. As the aforementioned idea of backup topology generation, we try to reduce the load of the controller by locally restoring the failed connection. The effectiveness of this approach is measured through the metric below.
\begin{definition}
\textbf{(Controller Overhead)}
Controller overhead is defined as the proportion of reconnection requests that need to be processed by the controller.
\end{definition}
\item \emph{Maximal Load}. As defined in Section \ref{sec:splicing-module}, it quantifies the maximal load among nodes after a regional failure.
\end{itemize}

We compare our SDN-based Fast and Resilient Routing against Disaster (SDN-FRRD) approach with the following approaches proposed in the literature,
 \begin{itemize}
\item MRC\cite{kvalbein2006fast}. The MRC is designed for local fast recovery. We evaluate it to see if it's sufficient for regional failures.
\item SDN-MRC. We apply the MRC to our novel framework, by directly using the MRC in the Backup Topologies Generation module.
\item Path Splicing\cite{motiwala2008path}. The advanced multipath routing algorithm, path splicing, is to random splicing routes in the data plane. The setting of Path Splicing is: using the same number of $k$ backup topologies as MRC and SDN-MRC, the link weight perturbation function is: $weight(i,j)=(degree(i)+degree(j))/degree_{max}$ where $degree_{max}$ is the maximal node degree and $weight(i,j)$ ranges from $0$ to $2$.
\end{itemize}
\subsection{Evaluation Results}
\begin{figure*}[!htb]
  \centering
  \subfigure[Germany backbone]{
    \label{50:result2} 
    \includegraphics[width=2.0 in]{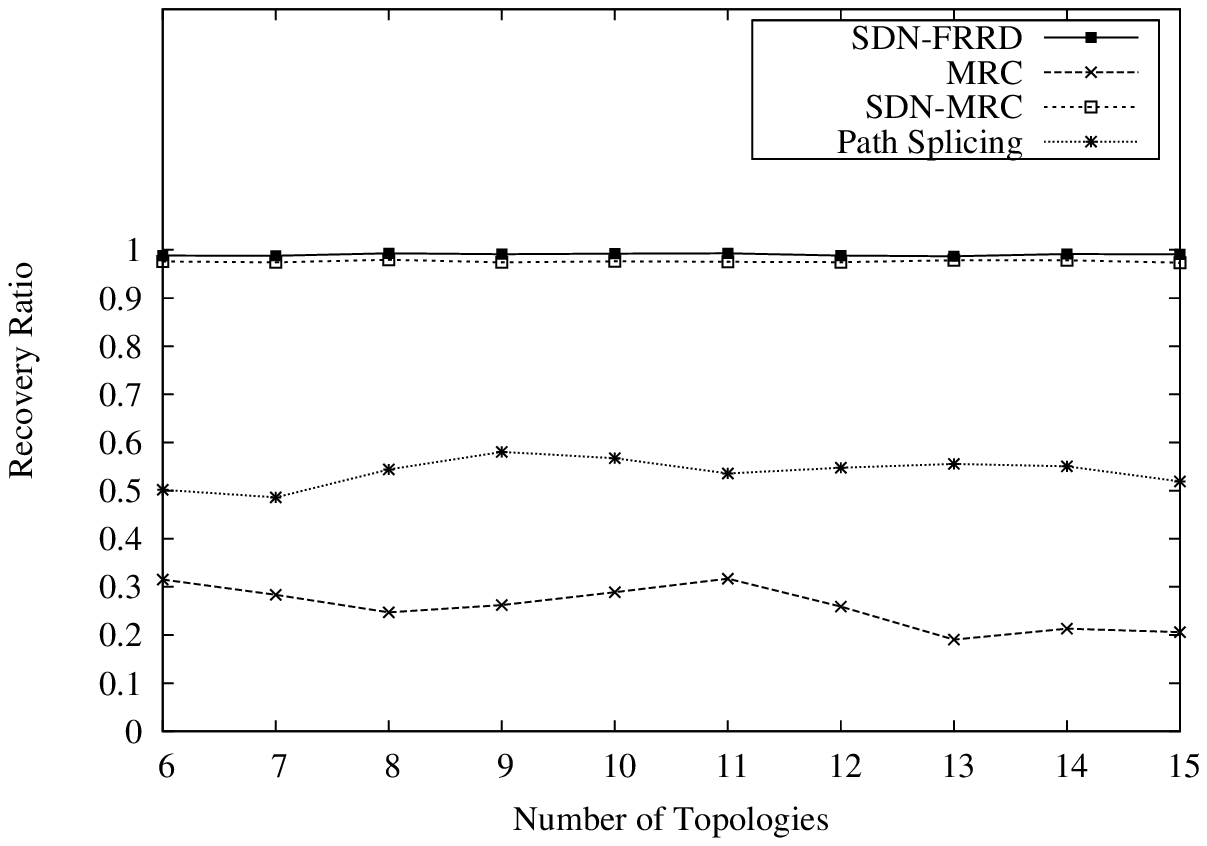}}
     \subfigure[Rand50]{
    \label{50:result3} 
    \includegraphics[width=2.0 in]{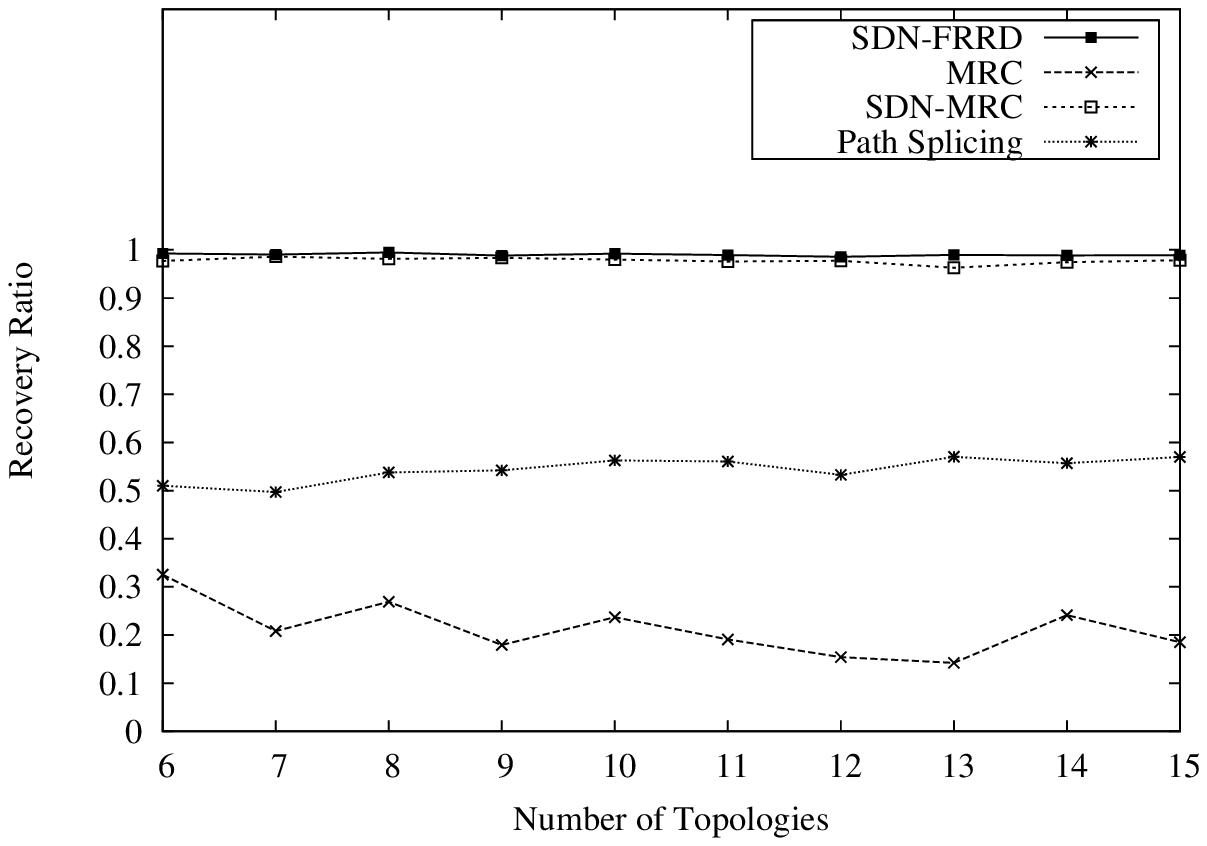}}
    \subfigure[Rand100]{
    \label{50:result3} 
    \includegraphics[width=2.0 in]{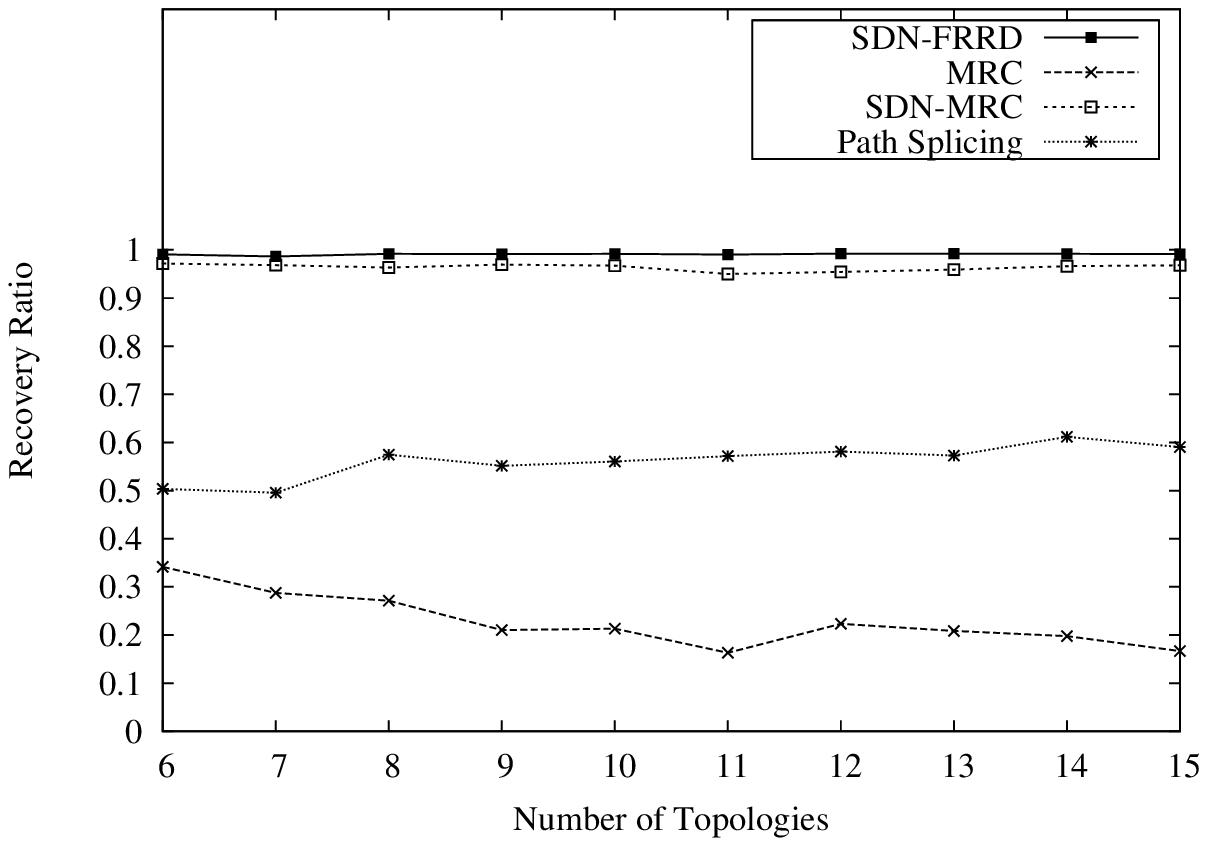}}
      \caption{Recovery Ratio vs. Number of backup topologies $k$ when the failure radius=50}
  \label{fig:recovery_result_radius_50} 
\end{figure*}

\begin{figure*}[!htb]
  \centering
  \subfigure[Germany backbone]{
    \label{100:result2} 
    \includegraphics[width=2.0 in]{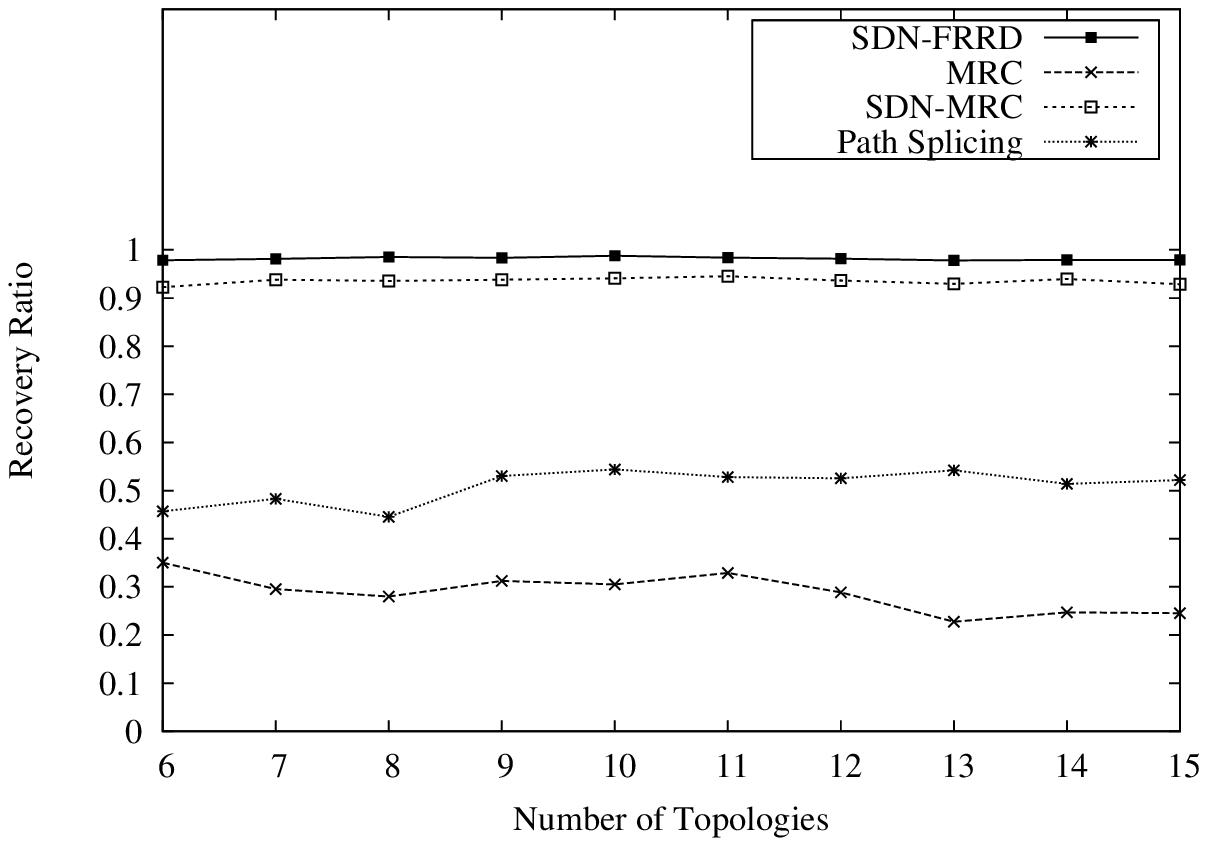}}
     \subfigure[Rand50]{
    \label{100:result3} 
    \includegraphics[width=2.0 in]{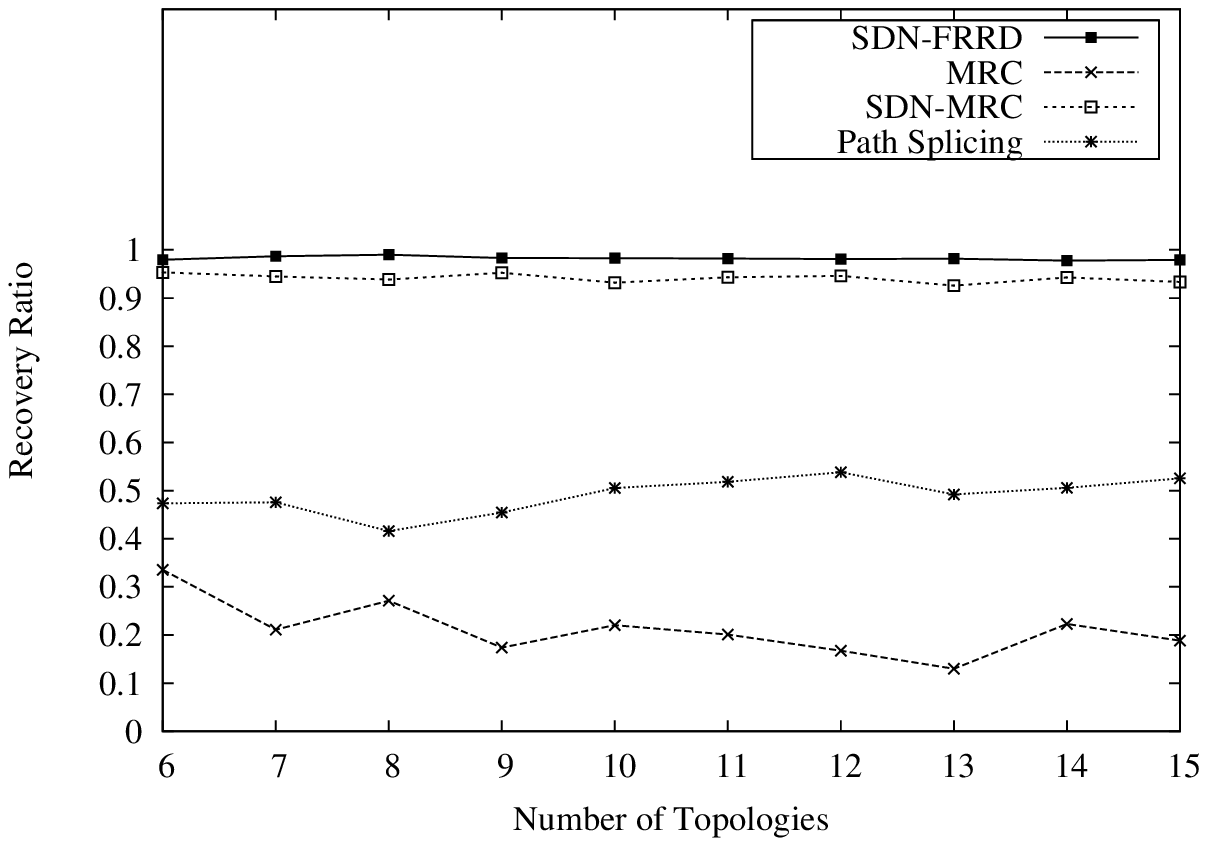}}
    \subfigure[Rand100]{
    \label{100:result3} 
    \includegraphics[width=2.0 in]{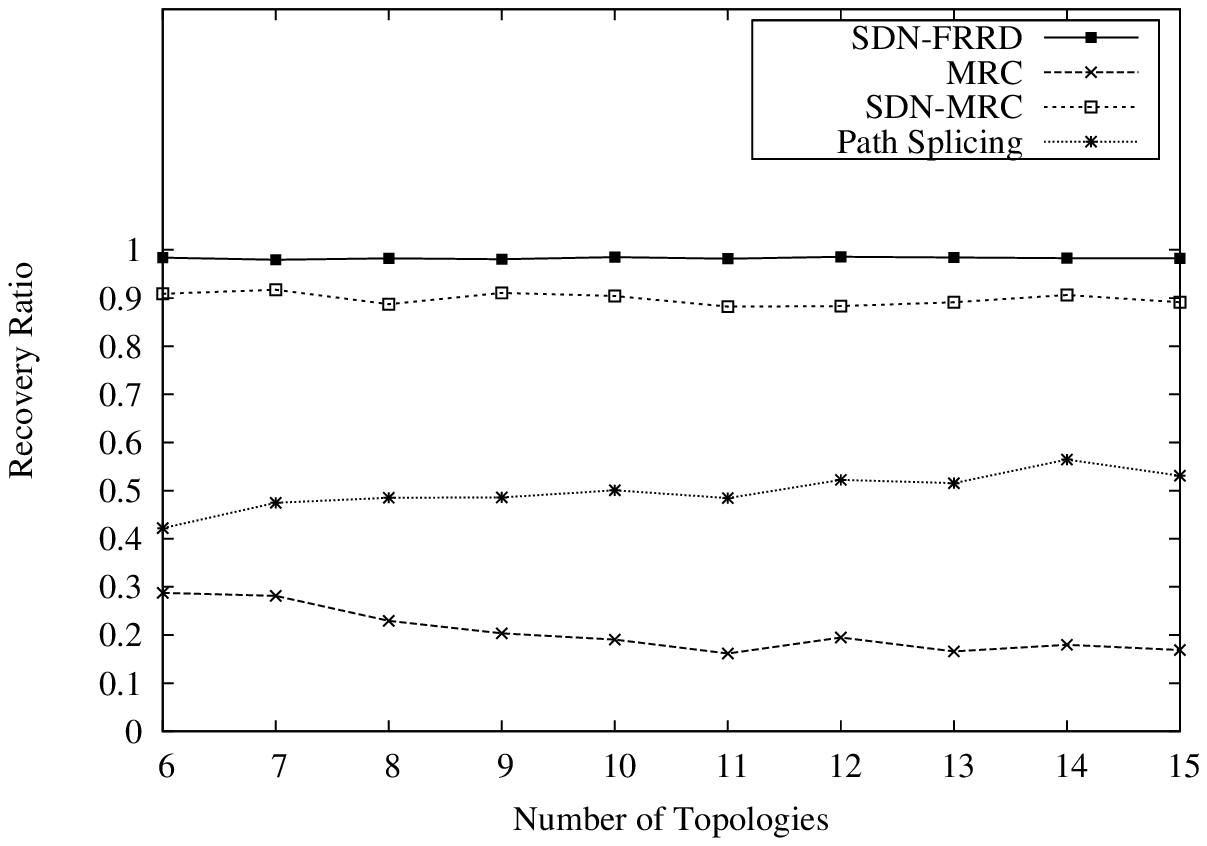}}
      \caption{Recovery Ratio vs. Number of backup topologies $k$ when the failure radius=100}
  \label{fig:recovery_result_radius_100} 
\end{figure*}

\begin{figure*}[!htb]
  \centering
  \subfigure[Germany backbone]{
    \label{stretch:result2} 
    \includegraphics[width=2.0 in]{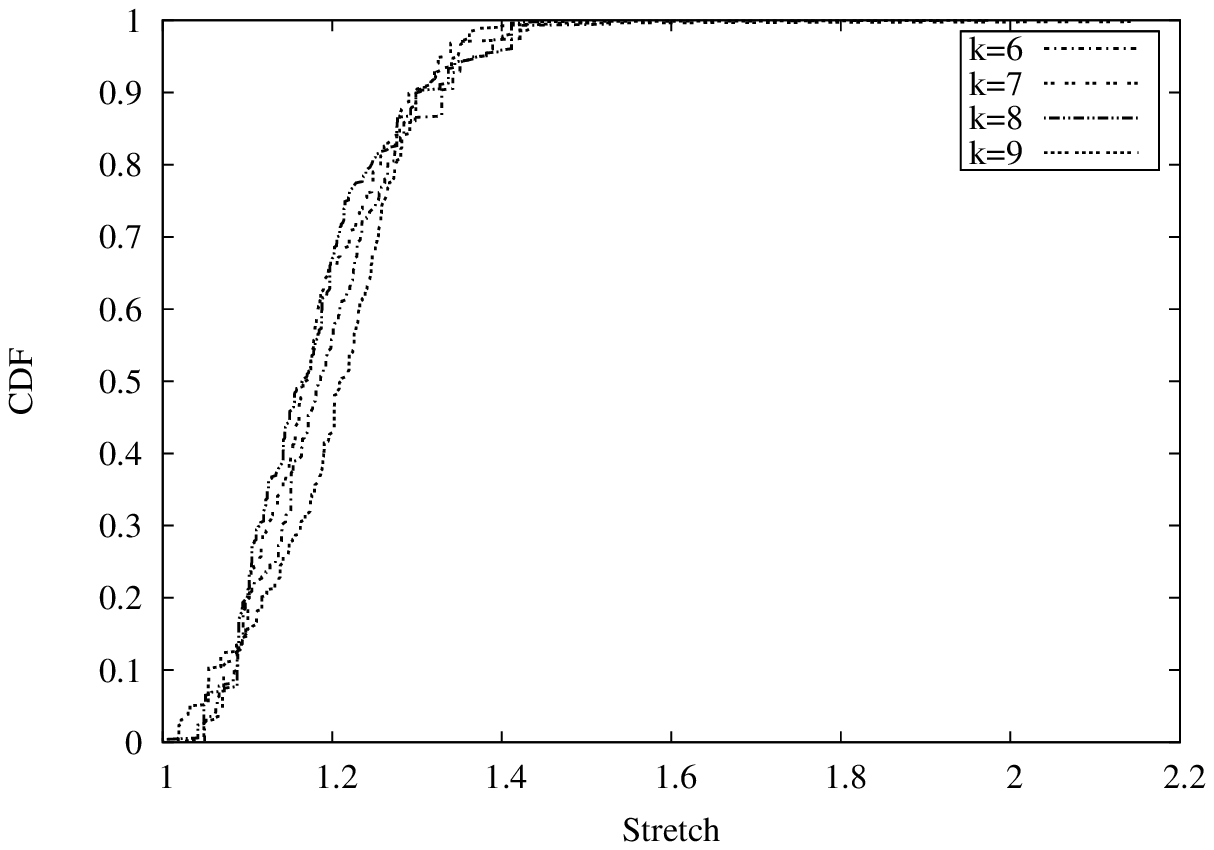}}
     \subfigure[Rand50]{
    \label{stretch:result3} 
    \includegraphics[width=2.0 in]{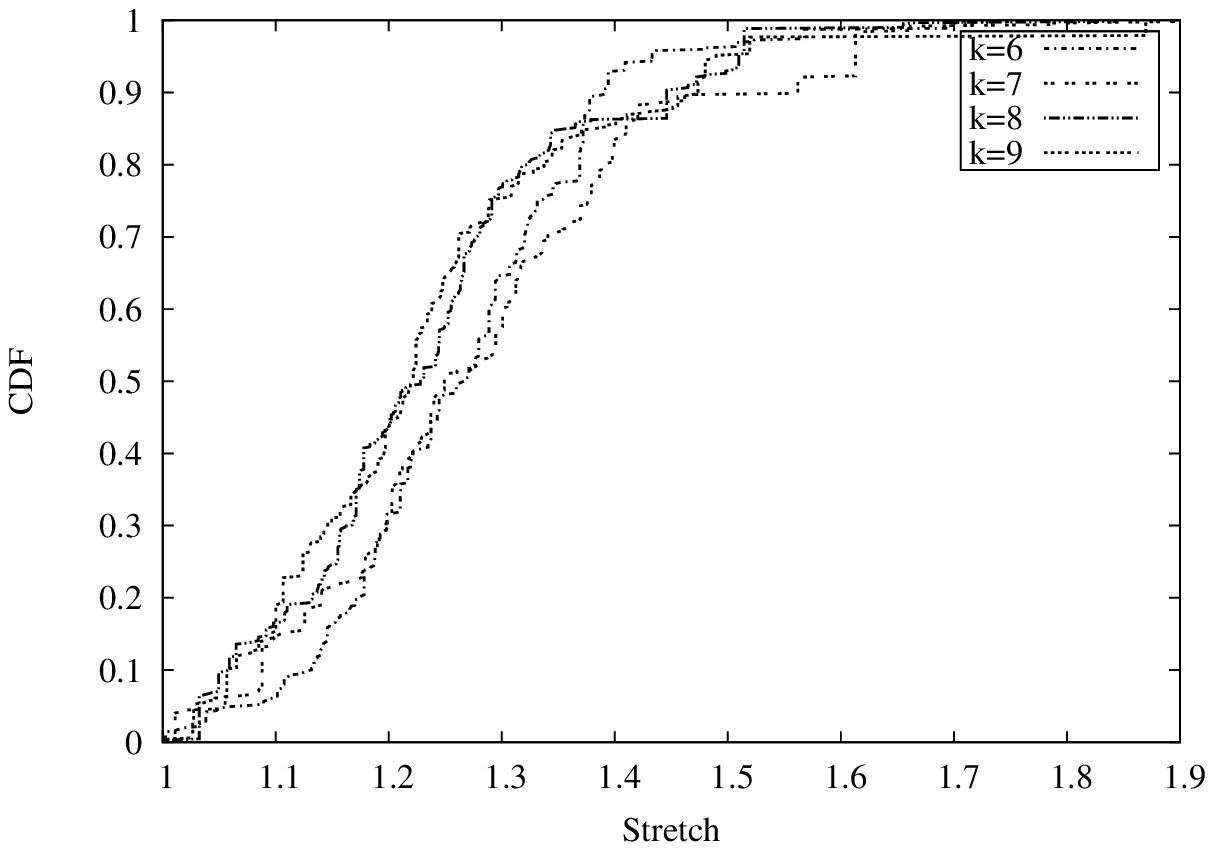}}
    \subfigure[Rand100]{
    \label{stretch:result3} 
    \includegraphics[width=2.0 in]{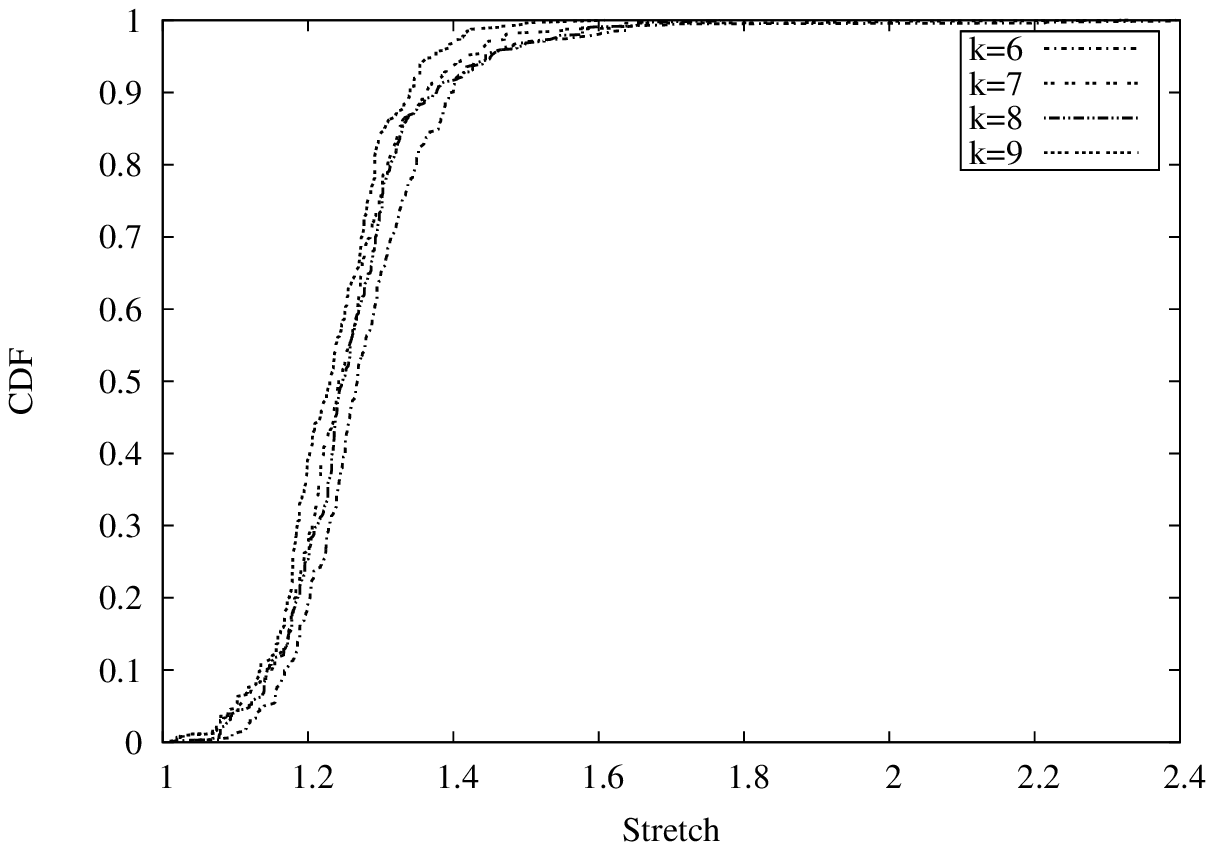}}
      \caption{Path Stretch vs. Number of backup topologies $k$ when the failure radius=50}
  \label{fig:stretch_result_radius_50} 
\end{figure*}

\begin{figure*}
\begin{minipage}[t]{0.23\linewidth}
\centering
\includegraphics[width=1.8 in]{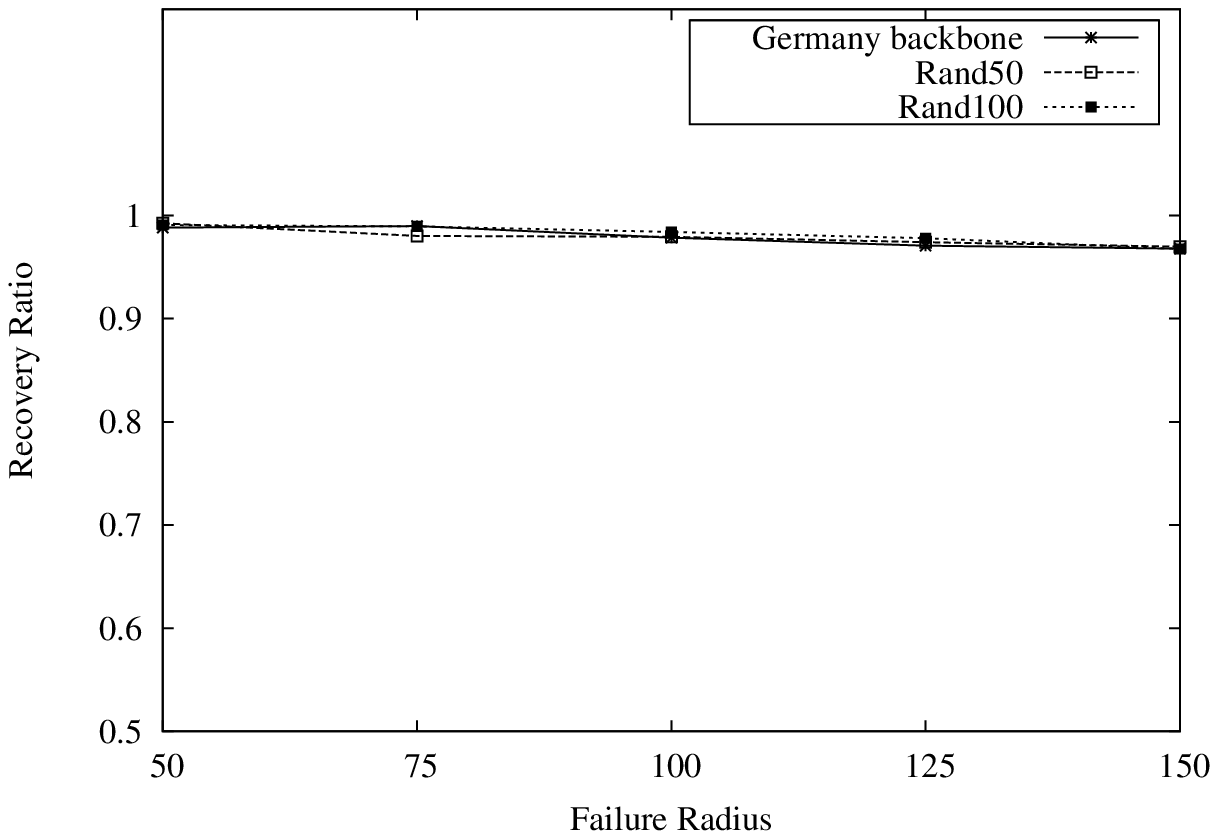}
\caption{Recovery Ratio vs. Failure Radius}
\label{fig:recovery-rand100-asc}
\end{minipage}\hfill
\begin{minipage}[t]{0.23\linewidth}
\centering
\includegraphics[width=1.8 in]{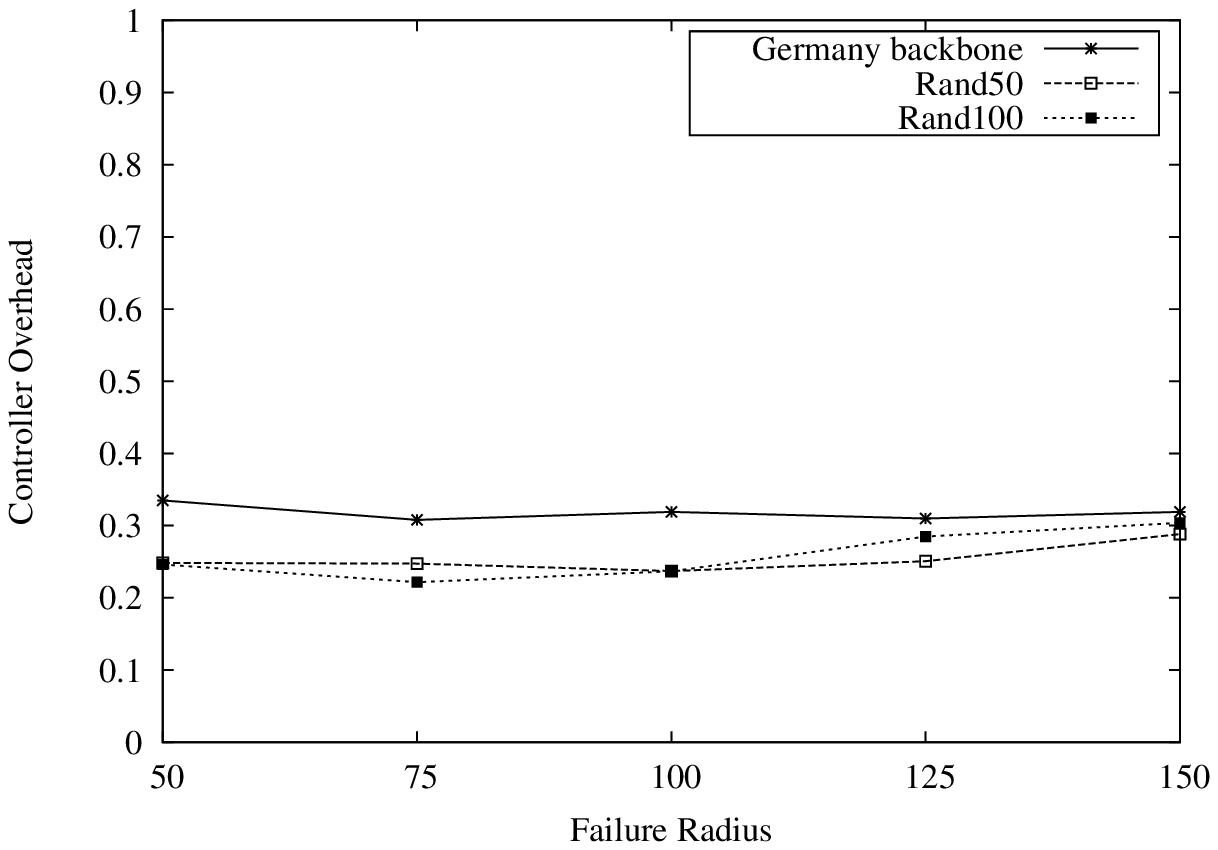}
\caption{Controller Overhead}
\label{fig:controller-overhead}
\end{minipage}\hfill
\begin{minipage}[t]{0.23\linewidth}
\centering
\includegraphics[width=1.8 in]{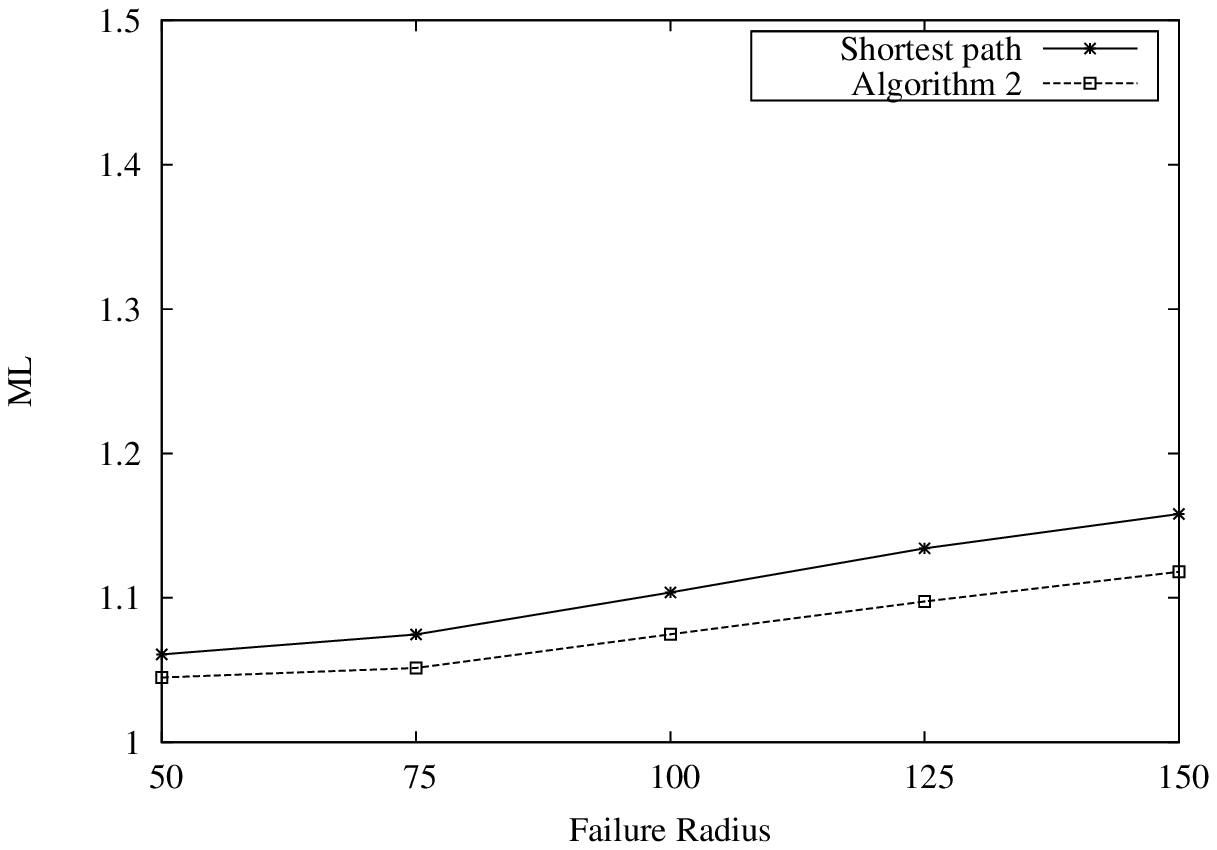}
\caption{ML on the Germany backbone}
\label{fig:ufd}
\end{minipage}\hfill
\begin{minipage}[t]{0.23\linewidth}
\centering
\includegraphics[width=1.8 in]{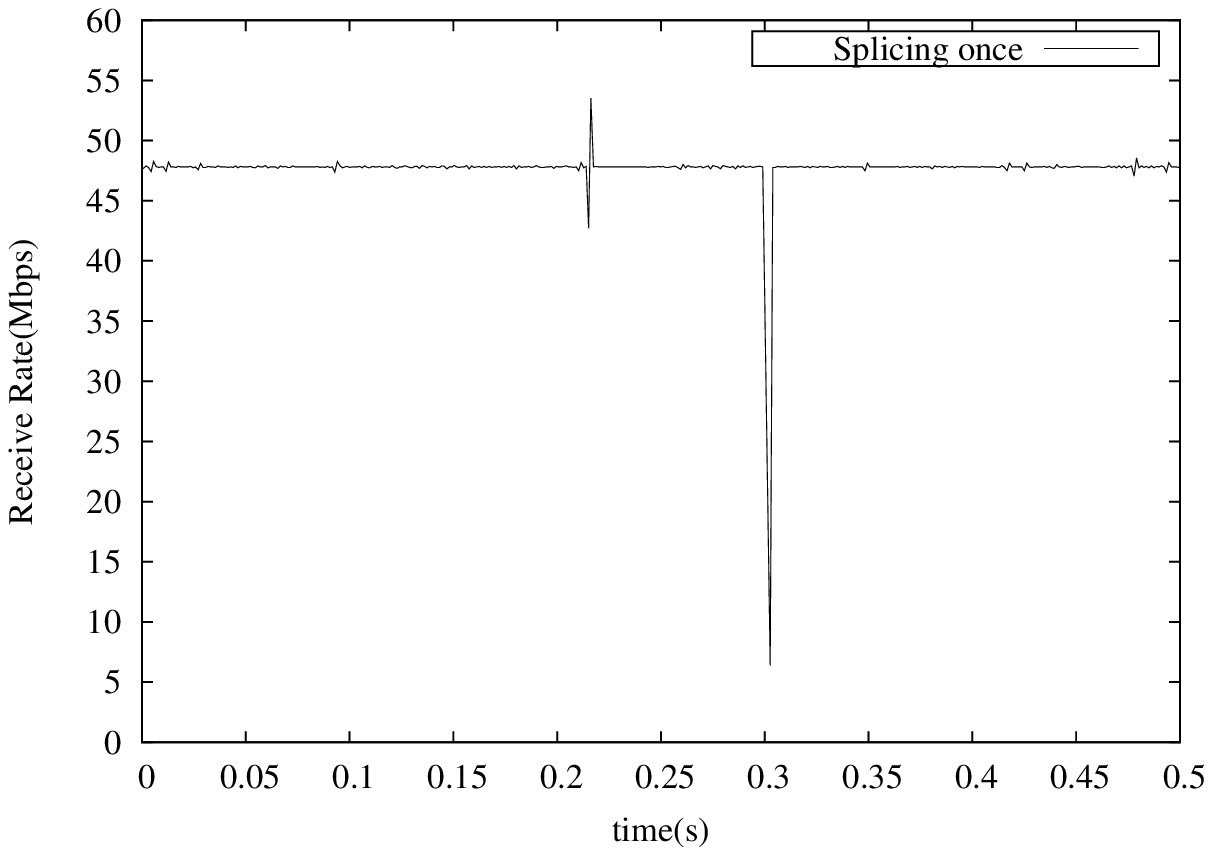}
\caption{Receive Rate on the Iperf client}
\label{fig:bandwidth_1}
\end{minipage}
\end{figure*}
\subsubsection{Recovery Ratio}
Fig. \ref{fig:recovery_result_radius_50} shows the recovery ratio in term of the number of backup topologies $k$ in the three topologies when the radius of the regional damage is $50$. From the graph, we can see that both SDN-FRRD and SDN-MRC, that apply the SDN framework can steadily achieve more than $90\%$ recovery ratio. Comparing to the MRC and the Path Splicing curves, clearly shows the effectiveness of our SDN framework.

Fig. \ref{fig:recovery_result_radius_100} shows the recovery ratio when the radius of the regional failure is $100$. The trend of the curves is similar to the ones inFig. \ref{fig:recovery_result_radius_50}. When the failure radius is $100$, the failure breaks more links than when the failure radius is 50. Thus the recovery ratio of the MRC, SDN-MRC, Path Splicing gets decreased. For example, the recovery ratio decrease by about $5\%$ in Fig. \ref{100:result2} compared to Fig. \ref{50:result2}, and decreases by about $10\%$ in Fig. \ref{100:result3} compared to Fig. \ref{50:result3}. However, we observe no significant decrease of the curve SDN-FRRD. This is because in the Backup Topologies Generation module (see Algorithm \ref{algo:b_t_g}), we adopt the vulnerable area of a path and consider the distribution of the failure radius to generate backup routes, such that the recovery ratio is not significantly influence by the size of the regional failure. This can also be validated in Fig. \ref{fig:recovery-rand100-asc}, where the recovery ratio remains above $95\%$ even when the failure radius is $150$ in all the three topologies.

Since the recovery ratio of the SDN-FRRD is almost about $100\%$ and is steady when the number of the backup topologies $k$ is from $6$ to $15$. Since small $k$ already has satisfying performance of the recovery ratio, small values of $k$ is sufficient. Because larger $k$ means the backup tables would consume more switch resources, network operators who have a strict limitation of switch resources can consider choosing the smallest $k$.
\subsubsection{Stretch}
Fig. \ref{fig:stretch_result_radius_50} shows the stretch in term of $k=6, 7, 8, 9$ in the three topologies. As we can see, in all the three topologies, about $90\%$ of the stretch is below 1.5. Normally, larger $k$ means more redundancies, which can lead to smaller stretches. The four values of $k$ achieve approximately equal recovery ratio, larger $k$ tends to have smaller stretch. This can be seen, for example, in Fig. \ref{stretch:result3}, the curve of $k=9$ is on the left side of the curve of $k=6$, which means a smaller stretch. This leads to a trade off between cost and performance at the initialization of network, i.e., operators who want to get a lower stretch can choose larger $k$, at the cost of more switch/router routing tables consumptions.
\subsubsection{Controller Overhead}
Fig. \ref{fig:controller-overhead} shows that when $k=6$, the controller overhead in terms of the failure radius. In all the radiuses, the controller only needs to handle about $40\%$ of failures, which means the data plane has already handle more than $60\%$ of the failures. As the failure radius grows bigger, the controller overhead has the trend to get heavier too. This is because when more links are destroyed, it is more difficult for the data plane to recover from the failure. Even when the failure radius is 150, about $60\%$ are handled locally.
\subsubsection{Maximal Load}
Fig. \ref{fig:ufd} shows the maximal load of the splicing actions generation algorithm (Algorithm \ref{algo:s_a_g}), compared to the shortest path splicing actions generation. The shortest path splicing actions generation choose the path with the minimal path length between a reconnection request $(s, t)$ when multiple rerouting paths between them are available\cite{xie2014designing}. It however does not consider the load distribution among nodes. The results of the ML reduction are normalized based on the result of ILP1. From the graph, we can see that our algorithm can reduce the ML. As the failure radius becomes bigger, the ML also gets bigger, which indicates that without consider the load distribution, the load imbalance among node gets more severer.
\subsubsection{Recovery Time}
Fig. \ref{fig:bandwidth_1} shows the receive rate on the Iperf client. A region failure occurred between the Iperf server and client at 0.3s. Packets can not be handled locally by backup tables, thus are sent to the controller. The receive rate on the client did have a sharp reduction at 0.3s, but it recovered very fast after about 10ms. The recovery time in real scenarios differs, which depends largely on the round trip time between a switch and a controller.
\section{Related Work}
There are limited number of recent papers focusing on leveraging SDN for large-scale regional failures. Nguyen \textit{et al.} \cite{nguyen2013software} studied latency between a switch and a controller and confirmed the applicability of SDN on disaster-resilient WANs. Works in \cite{sharma2012openflow,staessens2011software} studied using SDN to meet carrier-grade requirements and pointed out that the reactive approach may not be able to achieve sub-50ms recovery. However, the above works did not consider the heavy controller overhead and the consistent packet processing problem\cite{katta2013incremental,peresini2013cpp}. To reduce the recovery time, Sgambelluri \textit{et al.} \cite{sgambelluri2013openflow} proposed the proactive segment protection. Kamamura \textit{et al.} \cite{kamamura2013autonomous} gave a prototype to achieve IP fast rerouting using backup tables via autonomous OpenFlow controllers. The proposed proactive recovery can significantly reduce the recovery time. But, in face of region failure scenarios, the performance may be significantly decreased since the flexibility of SDN's global view is not used.
\section{Conclusion}\label{sec:conclusion}
In this paper, we propose a SDN based architecture to enhance the reliability of network against disaster failures. We propose our algorithms for geographic-based backup topologies generation and splicing considering the laod distribution among nodes, and implement our approach by utilizing multiple tables pipeline processing and fast failover group tables of OpenFlow. Experiments show that, by well pre-designed backup topologies protection, our fast restoration approach can efficiently use the redundancy to achieve high reachability and low stretch with low controller overhead. The load distribution after a regional is more even, compared to the previous splicing algorithm in \cite{xie2014designing}.

\bibliographystyle{IEEEtran}
\bibliography{tech-report}

\end{document}